\documentclass{vgtc}                          

\ifpdf
  \pdfoutput=1\relax                   
  \usepackage{graphicx}                
  \DeclareGraphicsExtensions{.pdf,.png,.jpg,.jpeg} 
\else
  \usepackage{graphicx}                
  \DeclareGraphicsExtensions{.eps}     
\fi%

\graphicspath{{figures/}{pictures/}{images/}{./}} 

\usepackage{microtype}                 
\PassOptionsToPackage{warn}{textcomp}  
\usepackage{textcomp}                  
\usepackage{mathptmx}                  
\usepackage{times}                     
\usepackage{cite}                      
\usepackage{tabu}                      
\usepackage{booktabs}                  
\usepackage{float}

\onlineid{0}

\vgtccategory{Research}

\vgtcinsertpkg

\title{Brown Ring Experiment in Virtual Reality}

\author{Prithaj Jana\thanks{e-mail: jprithaj@cs.stonybrook.edu}\\ %
        \scriptsize Stonybrook University %
\and Emil Joswin\thanks{e-mail: ejoswin@cs.stonybrook.edu}\\ %
     \scriptsize Stonybrook University}

\abstract{Brown Ring Experiment is a very popular test to detect the presence of Nitrate in salts commonly performed in chemical laboratories with supplies of required chemicals. Our work clears out the need for a chemical laboratory and chemicals in order to understand the experiment practically. We have used the technology of Virtual Reality to fulfill this requirement. Our research work can be extensively utilized to create virtual environments for conducting other chemical processes in a virtual environment hence, eliminating the need for a chemical laboratory. This can help students in remote areas with minimal resources to fill in the void of
practical experiments they have in their learning process due to space constraints.%
} 

\CCScatlist{
  \CCScatTwelve{Virtual Reality}{Unity3D}{HTC Vive}{Education};
  \CCScatTwelve{Chemistry}{Virtual Worlds}{}
}

\begin{document}


\firstsection{Introduction}

\maketitle

There is a considerable crunch in the availability of Chemical
laboratories in remote parts. So, students lack the resources to gain
practical knowledge of the concepts they learn in class. So, our
initiative is to provide that practical experience using virtual reality
so that students don’t require the presence of an actual chemical
laboratory to gain practical knowledge. For demonstrating this, we
have taken up a very popular chemical experiment to detect the
presence of Nitrate in salts called Brown Ring Experiment.

To start with, let’s see what wikipedia has to say about the Brown
Ring Experiment. Wikipedia says, A common nitrate test, known as
the brown ring test \cite{ref1} can be performed by adding iron(II) sulfate to
a solution of a nitrate, then slowly adding concentrated sulfuric
acid such that the acid forms a layer below the aqueous solution. A
brown ring will form at the junction of the two layers, indicating the
presence of the nitrate ion.\cite{ref2} Note that the presence of nitrite ions
will interfere with this test.\cite{ref3}

The overall reaction is the reduction of the nitrate ion by iron(II)
which is oxidised to iron(III) and formation of a nitrosonium
complex where nitric oxide is reduced to NO$^-$.\cite{ref4}
\\
\\
2HNO$_3$+3H$_2$SO$_4$+6FeSO$_4$ --$>>$ 3Fe$_2$(SO$_4$)$_3$+2NO+4H$_2$O \\
(Remaining)[Fe(H$_2$O)$_6$]SO$_4$+NO=[Fe(H$_2$O)$_5$(NO)]SO$_4$+H$_2$O
\\
\\
This test is sensitive up to 2.5 micrograms and a concentration of 1
in 25,000 parts.\cite{ref5}

This article will demonstrate a way to set up equipments and
chemicals needed to conduct this experiment on to a scene in the
virtual environment and the user can conduct the experiment by
picking up stuff and dropping them according the demands of the
experiment.

\section{Related Work}
A large number of researchers have studied the impact of virtual reality technologies in chemical education because it is believed that students can form appropriate mental models of a concept by visualizing and interacting with the representation of the phenomenon.
In 2006, Kerawalla, Luckin, Seljeflot \& Woolard\cite{ref6} have suggested
four design requirements that need to be considered for AR to be
successfully adopted into classroom practice in the future. These
requirements according to the authors are: Flexible content that
teachers can adapt to the needs of their children; guided.
exploration so learning opportunities can be maximized in a limited
time; and attention to the needs of institutional and curricular
requirements. In 2008, Maria Limniou, David Roberts \& Nikos Papadopoulos \cite{ref7}, also demonstrated the use of CAVE for a fully immersive virtual environment for chemistry education. In 2011, a paper by L.D. Antonogluu, N.D. Charistos \& M.P. Sigalas \cite{ref8} implemented a design for an enhanced hybrid course on molecular symmetry. In 2013, a paper by P. Maier and G. Klinker\cite{ref9} also
used Augmented Reality to support Chemistry students in learning
and researchers in developing and understanding new chemical
molecular structures and their spatial relations. In 2014, Su Kai, Xu Wang \& Feng-Kuang Chiang \cite{ref10} demonstrated an augmented reality based simulation system for visually understanding chemical reactions using markers. 

\section{Method}
Our method comprises carrying out chemical experiments by mixing
virtual chemicals. We use VR device controllers to move the
equipments, containing chemicals in chemistry lab. In this
experiment we have glass bottles comprising the Ferrous Sulphate
solution, Sulphuric acid and Nitrate solution. The glass bottle
consisting of Ferrous Sulphate is picked using the HTC Vive
controller and its contents are passed into the test-tube by tilting the
glass bottle above the mouth of the test tube. Then we add the Nitrate
solution (salt) the same way. After this, the dropper is picked up with
the controller and placed inside the glass bottle containing Sulphuric
acid. Small quantity of Sulphuric acid is sucked in by the dropper.
This is later dropped into the same test tube through the sides of the
test tube. On contact with the Sulphuric acid particles, the mixture
undergoes a chemical reaction and a thin brown ring is formed
around the test-tube slightly above the middle point of the solution.

\section{Implementation Details}
The liquid particles were simulated via transparent spheres. Models
for the experiment were designed by ourselves using the 3D
modelling/sculpting software Maya. The table upon which the
experiments are setup is downloaded for free from free3d.com\cite{ref11}.
The glass bottles had a mesh collider in the inside to prevent the
particles from falling through. Each of the particles had their own
sphere collider attached to themselves and hence, they were stuck
within the mesh. It was also required to have a collider for the HTC
Vive controller to grab onto. This was achieved by creating an empty object collider as a sibling to the original glass bottle. They are then
given a common parent and this parent gets attached to the controller
when the controller grabs the collider. This way we can grab and
move the entire bottle and its contents towards the test tube\cite{ref12} or any
other position desired.

The test tube contains a collider on the inside that would prevent
the objects from moving out of it once fallen inside so they are not
impacted by the movement of the object they have fallen from.
While the particles are being transferred from the glass bottle into
the test tube, we are changing the parent of the particle from the
glass bottle to the test tube to attain that functionality. This is done
only for the particles that have fallen into the test tube. During the
pouring process, we calculate the distance between the particle and
the test tube and the glass bottle. If the test tube is within a
threshold, then the test tube becomes the parent of the particle. The
same check is performed between the particle and the glass bottle. If
there are particles that are outside the threshold for both of them,
then those are the particles that must’ve slipped though the edges of
the test-tube while pouring. For such particles we set their parent to
be null.

      \begin{figure}[h]
        \center{\includegraphics[width=240px]
        {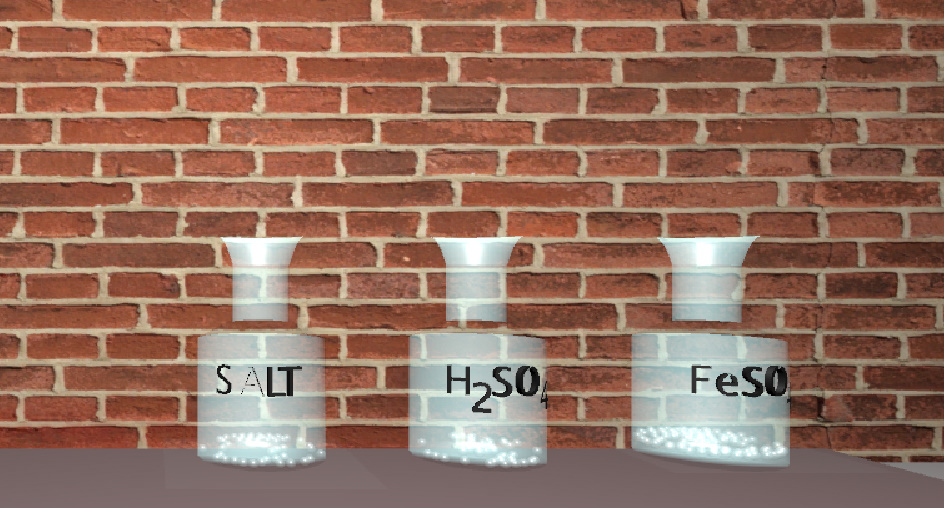}}
        \caption{\label{fig:my-label} Glass bottles with Chemicals}
      \end{figure}

The dropper pipette is designed to have particle attractor in the
interior that gets activated only upon pressing the dropper’s bulb at
the top. For this we have invisible plane at the mouth of the dropper
that open and closes upon the press of a button. Once this plane is
open, the contents inside the dropper falls out due to gravity. They
then become the child of the object they are falling into. If they fall
on the ground, their parent becomes null. Once inside a glass bottle,
we open the mouth of the dropper and enable suction of the particles
by transforming the particles into the magnet that is contained inside
the dropper.

The mouth closes when sufficient number of particles goes in.
The particles that are transferred into the dropper, then drops
towards the mouth of the dropper and stays there until the mouth is
opened.

      \begin{figure}[H]
        \center{\includegraphics[width=80px]
        {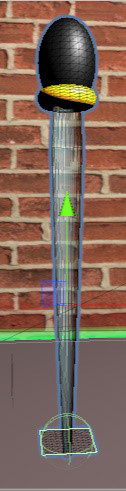}}
        \caption{\label{fig:my-label} Dropper Pipette}
      \end{figure}

Once the contents inside the dropper pipette is dropped into the
test tube, the reaction commences. A brown ring is formed inside the
test tube about 4/5th of the way from the bottom of the test tube. This is accomplished by first figuring out the total height of the particles
inside the test tube along the axis of the test tube and dividing them
into 5 and picking the 4th layer. We then colour all the particles in
this layer that are closer to the test tube. 

      \begin{figure}[H]
        \center{\includegraphics[width=240px]
        {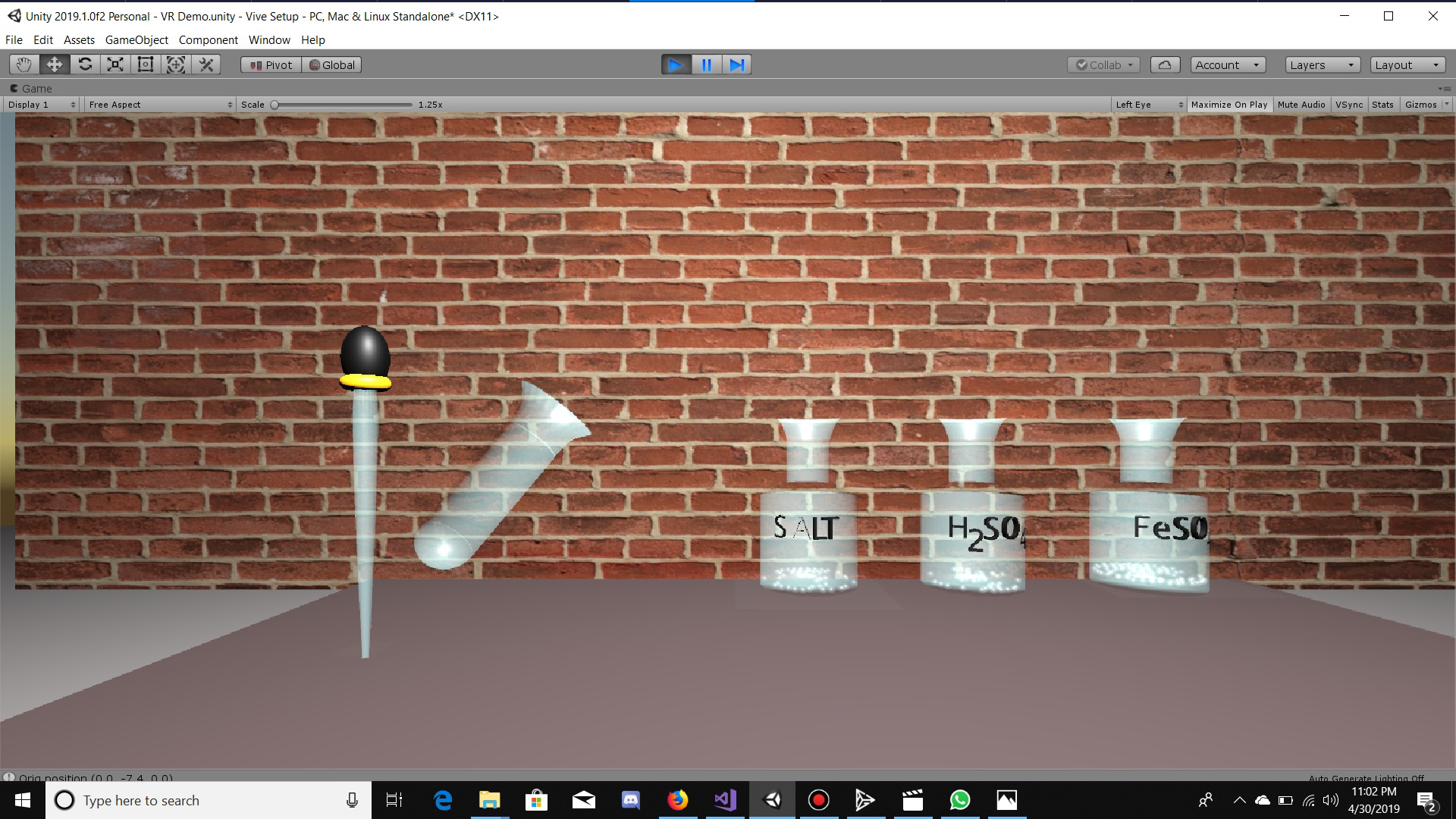}}
        \caption{\label{fig:my-label} Full Scene}
      \end{figure}

One thing that has to be
maintained is the order of the chemicals when added to the test-tube.
At first iron sulphate must go in, followed by salt solution and lastly
the concentrated sulphuric acid is to be added by the dropper
through the sides to see the results.

      \begin{figure}[H]
        \center{\includegraphics[width=240px]
        {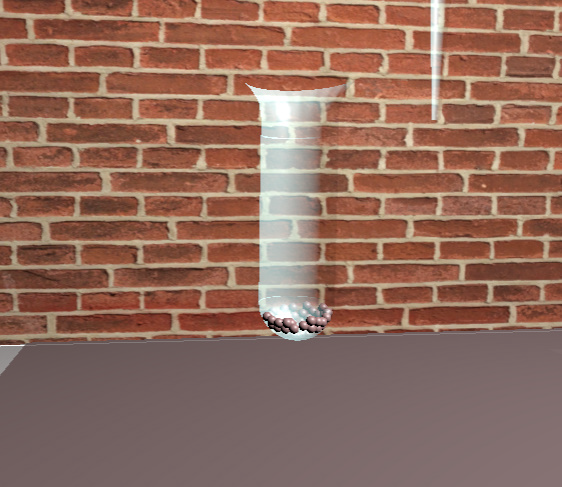}}
        \caption{\label{fig:my-label} Final Result - Brown Ring Formation}
      \end{figure}

We can hence see the results in Figure 4 with the brown ring formed to complete the experiment.

\section{Results}
The experiment was completed successfully except for the
functioning of the spheres inside equipments imitating liquids. When the glass bottles are moved fast,
the contents inside sometimes tend to fall off. This happens only
when the controller is used. However, this cannot be reproduced if
the glass bottles are moved using keyboard controls.
Another issue associated with the same bug is that once a particle.
move outside the glass bottle, it stays there indefinitely. Our best
guess for the cause would be an improper modelling of the equipment mesh in Maya due to our limited exposure to the software. As there are imperfections in the model the spheres find gaps to get out of the equipment mesh and hence, the fall off and the indefinite stay outside the mesh. These errors can be corrected by getting the models prepared by experienced Maya sculptors.

\section{Conclusion}

We have demonstrated here that you can perform chemical
experiments in a virtual reality environment successfully. Further
experiments can be conducted by adding new scenes and creating a
user control on which experiment the user want.
The boundaries of this observation is not limited to Chemistry
alone and can be used to successfully conduct experiments in other
subjects also. Additionally, a cloud-based platform can also be created to allow 
developers to add such creative educational content on a shared area 
to be used by educational instituations all around the globe.

\acknowledgments{
We thank Prof. Arie Kaufman from Department of Computer Science at Stony Brook University for allowing us to use the equipments in the VR lab to complete the demonstration of our project. We also thank Mr. Saeed Boor Boor, PhD Student and Ms. Ping Hu, PhD Student for their endless mentorship during the project work. I would like to thank my friends for accepting nothing less than excellence from me. Last but not the least, I would like to thank my family: my parents and to my brothers and sisters for supporting me spiritually throughout writing this paper.}

\bibliographystyle{abbrv-doi}

\end{document}